# β-$Ga_2O_3$ Proton Detection for Dosimetry in Intensity-Modulated Proton Therapy

Hunter D. Ellis[1, a)], Ajayvarman Mallapillai[1], Jared Miller[1], Imteaz Rahaman[1], Botong Li[1], Vikren Sarkar[2], and Kai Fu[1,a)]

[1]*Department of Electrical and Computer Engineering, The University of Utah, Salt Lake City, UT 84112, USA*

[2]*Department of Radiation Oncology, The University of Utah, Salt Lake City, UT 84112, USA*

**Abstract**

Intensity-modulated proton therapy (IMPT) employs proton radiation rather than conventional X-rays to treat cancerous tumors. This approach offers significant advantages by minimizing the radiation exposure of surrounding healthy tissue, leading to improved patient outcomes and reduced side effects compared to traditional X-ray therapy. To ensure patient safety, each treatment plan must be experimentally validated before clinical implementation. However, current dosimetry devices face limitations in performing angled beam measurements and obtaining multi-depth assessments, both of which are essential for verifying IMPT treatment plans. In this study, the performance of a β-$Ga_2O_3$-based metal–semiconductor–metal (MSM) detector with a low-noise amplifier has been studied and evaluated under various proton radiation doses and energy levels delivered by a MEVION S250i proton accelerator. The detector's performance was also compared with that of an ionization chamber. The β-$Ga_2O_3$ detector exhibited a linear response with proton dose for single-spot irradiations, and its response to varying proton energies closely matched both the ion chamber data and simulated dose distributions. These findings highlight the potential of β-$Ga_2O_3$-based detectors as robust dosimetry devices for IMPT applications.

[a)] Authors to whom correspondence should be addressed. Electronic mail: u0973796@utah.edu and kai.fu@utah.edu

## I. Introduction

Intensity-modulated proton therapy (IMPT) is a cancer treatment technique that reduces the risk of secondary malignancies and treatment side effects compared to conventional X-ray radiation therapy due to the Bragg peak in the energy deposition in the tissue[1-8]. Patient safety is ensured through an experimental verification of the radiation treatment plan, with the dosimetry commonly being done using ion chambers[9, 10]. However, ion chambers face difficulty in performing angled beam measurements, which is an essential measurement for IMPT plan verification, due to their high directionality[11]. Multi-depth measurements, another essential measurement for IMPT, can be performed using special stacked ion chamber arrays, but the price of these arrays limits their use in practice.

β-$Ga_2O_3$-based detectors are a promising alternative due to the material's high displacement energy and resulting radiation hardness[12, 13], which could allow for detectors that can perform angled beam measurements and multi-depth measurements. The radiation resistance of β-$Ga_2O_3$-based devices has already been successfully demonstrated, with devices maintaining functionality with proton doses between $5\times10^{1}$ $cm^{-2}$ to $10^{15}$ $cm^{-2}$ and over 1 Gy of gamma radiation[14-17]. This radiation hardness is comparable to other radiation-hard materials, such as 4H-SiC and GaN [18-20]. $Ga_2O_3$-based detectors have already been fabricated for various types of radiation, including alpha particles, neutrons, and X-rays[21-29], and they commonly benefit from high internal gain due to the low hole mobility of $Ga_2O_3$[30-33].

Previous studies have demonstrated the feasibility of β-$Ga_2O_3$-based metal–semiconductor–metal (MSM) detectors for proton radiation therapy applications, successfully detecting radiation across the entire repeatable dose range of a MEVION S250i accelerator [34]. However, these detectors exhibited elevated noise levels, and the effects of varying proton energy on the detector response were not thoroughly investigated. This work addresses these gaps by enhancing the signal fidelity of the β-$Ga_2O_3$ proton detector (GOPD) system, along with analyzing the effect of the proton beam energy on the response of the GOPD. The timing characteristics were also analyzed using single-spot proton radiation measurements with varying dose levels. The detector's performance was benchmarked against an ionization chamber, and its response across different proton energies was systematically analyzed.

## II. Methods

The GOPD was fabricated by the heteroepitaxial growth of β-$Ga_2O_3$ on sapphire using a Agnitron MOCVD system. The growth process occurred at 840 ℃, at a pressure of 60 Torr, and the process lasted 90 minutes. The shroud gas was set to a flow rate of 500 sccm, the $O_2$ flow rate was 800 sccm, the Ar flow rate was 1100 sccm, and the TEG precursors flow rate was set to 130 sccm. The growth resulted in a 1 μm thick β-Ga2O3 film with a donor concentration of approximately $7\times10^{15}$ $cm^{-3}$. The Schottky contacts were formed using lift-off to pattern 25 interdigitated metallic fingers, each measuring 490 μm in length and 10 μm in width, with a 10 μm spacing between adjacent fingers made from a Ni (60 nm)/Au (100 nm) bilayer. The fabricated device was subsequently mounted onto a printed circuit board, and electrical connections were established through wire bonding to enable signal readout.

An amplifier circuit was designed to process the detector output, and the schematic diagram is presented in Fig. 1(a). The circuit incorporated a charge-sensitive preamplifier that converted the detector's current pulse into a corresponding voltage signal. This output was then directed through a pulse-shaping filter, which filters out the noise while maintaining signal fidelity by utilizing a buffer amplifier. A final amplification stage is used to increase the gain of the voltage signal. The simulated gain of the complete preamplifier–amplifier configuration, along with the non-amplified total current signal from the GOPD to 40 MU of proton radiation at 200 MeV, is shown in Fig. 1(b). The amplifier had a max gain of 150 dB and a cut-off frequency of 8.8 kHz.

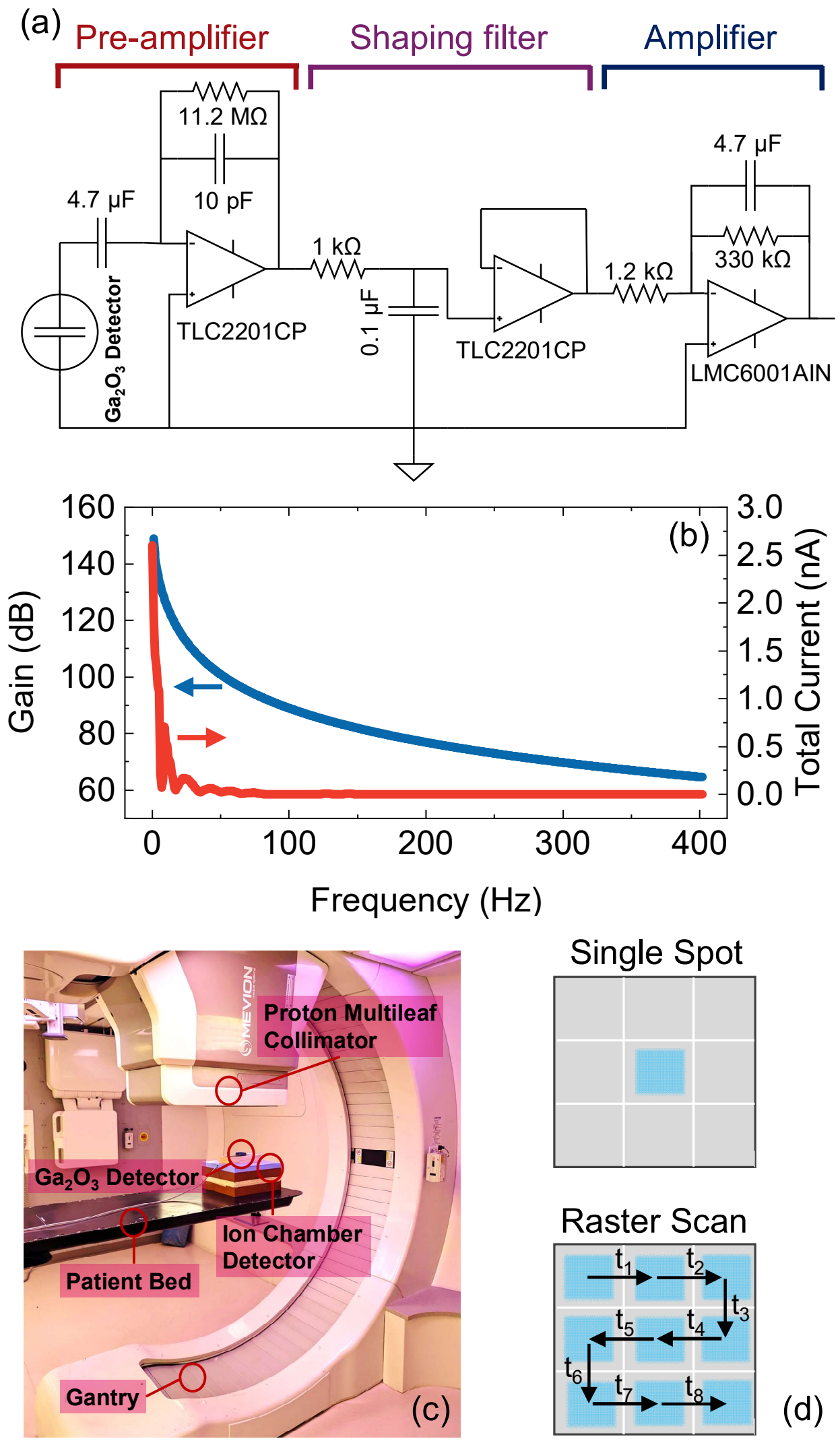

**Figure 1.** (a) Schematic of the preamplifier, signal shaper, and amplifier. (b) Simulated frequency response of the amplifier, along with the measured total current from the proton radiation detector to 40 MU and 200 MeV. (c) Experimental setup with the GOPD on top of the PPC05 plane-parallel ionization chamber, and directly below the proton emitter. (d) Schematic of single-spot and raster-scan radiation exposures.

For radiation testing, the GOPD was positioned beneath the proton radiation source on the treatment bed, as illustrated in Fig. 1(c). Proton irradiation was performed using a MEVION S250i accelerator system. The detector output voltage was monitored and recorded using an oscilloscope. To provide a reference for dose calibration, a PPC05 plane-parallel ionization chamber biased at 300 V was placed directly beneath the GOPD. Measurements were conducted for both single-spot proton exposures and raster-scan proton beams, the latter consisting of a series of spatially and temporally separated radiation spots. The schematic distinction between these irradiation modes is depicted in Fig. 1(d). The single-spot measurements were performed across a dose range of 0.26 MU to 16 MU and an energy range of 199 MeV to 60 MeV to characterize the detector and amplifier. The raster scan measurements were performed to emulate an exposure that a patient would receive, which were performed across a dose range of 0.26 MU to 4 MU at an energy of 190 MeV, with a 25 $cm^2$ area using 441 individual irradiation points.

## III. Results and Discussion

The performance of the GOPD was first characterized and evaluated using single proton pulses at 190 MeV while varying the delivered dose. A simulation of the deposited energy in bulk $Ga_2O_3$ is shown in Fig. 2(a), with a zoom in on a thickness equal to the GOPD. The dose levels ranged from 0.26 MU (3.0 mGy)—the minimum repeatable dose of the MEVION S250i system—up to 16 MU (0.185 Gy). For all measurements, the baseline voltage was offset to zero for clarity, and the resulting time-domain responses are displayed in Figs. 2(b)–2(e). Distinct output pulses were observed even at the lowest dose of 0.26 MU, with the pulse amplitude increasing systematically with higher doses. The preliminary voltage spike seen in Figs. 2(c)–2(e) corresponds to a position verification pulse from the accelerator, which is absent in the 0.26 MU case due to the low delivered dose. The rise time ($\tau_r$) was found by subtracting the time value of the voltage data at 10% of the peak value on the rising edge from the time value of the voltage data at 90% of the peak value on the rising edge. The fall time ($\tau_f$) was found using a decaying exponential fitting.

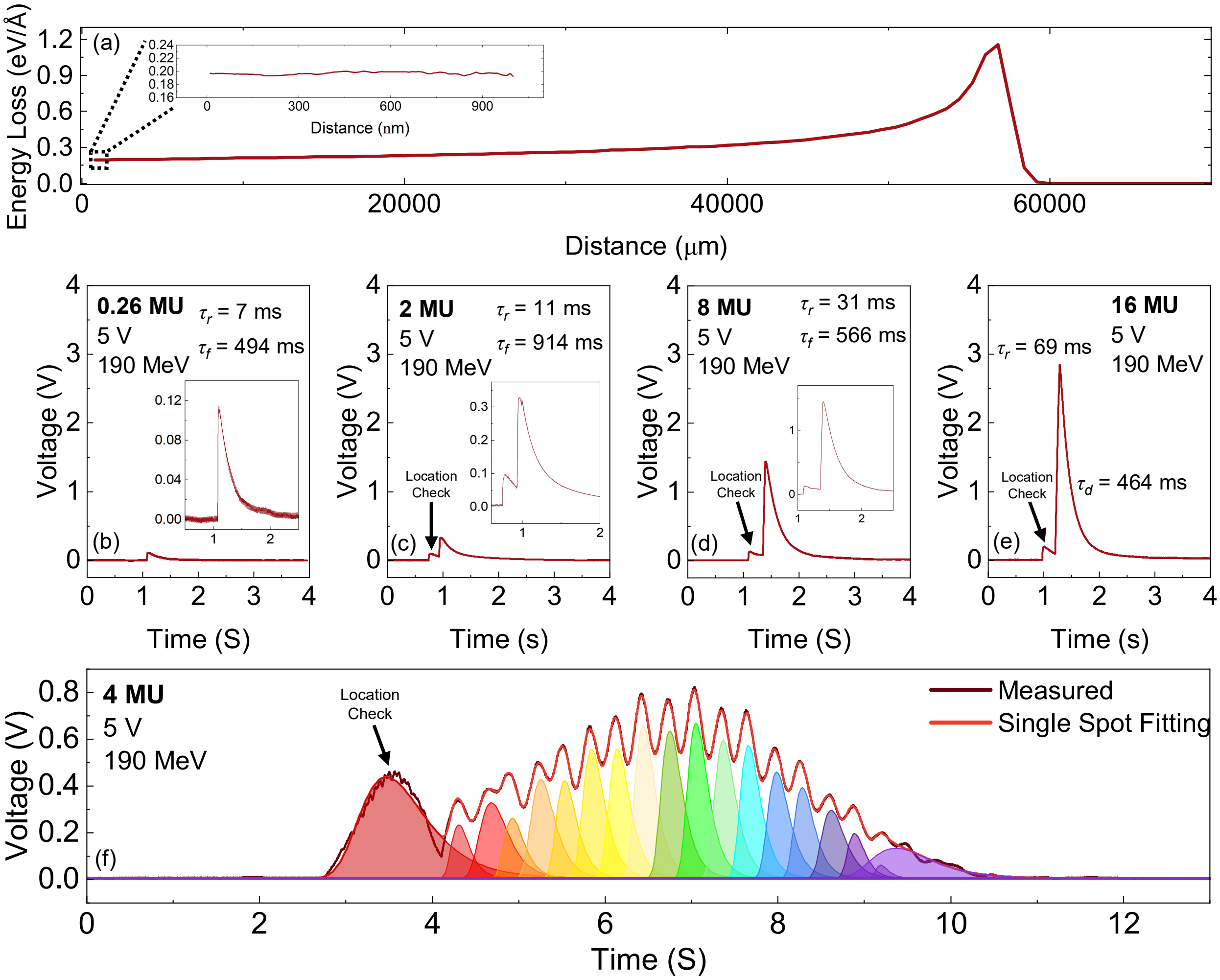

**Figure 2.** (a) Simulated energy loss versus distance. Transient voltage response of the GOPD and the amplifier with a single spot of proton radiation at 190 MeV with a dose of (b) 0.26 MU, (c) 2 MU, (d) 8 MU, and (e) 16 MU. The equivalent dose in Gy for 0.26 MU, 2 MU, 8 MU, and 16 MU is 3.0 mGy, 0.0231 Gy, 0.0923 Gy, and 0.185 Gy, respectively. The rise time ($\tau_r$) was calculated from the time difference of 90% of the peak value to 10% of the peak value on the rising edge. The fall time ($\tau_f$) was calculated using a decaying exponential fit. (f) Transient voltage response of the GOPD and the amplifier with a raster scan of proton radiation at 190 MeV and 4 MU per spot.

The transient response of the GOPD to a raster scan of proton radiation with a dose of 4 MU per spot at an energy of 190 MeV is presented in Fig. 2(f). Figure 2(f) reveals multiple voltage spikes corresponding to individual proton spots intersecting the GOPD, which has been displayed using a peak fitting. Each spot produced a distinct response separated in time, consistent with the spatial and temporal separation inherent to raster scanning.

To further analyze the detector response, the voltage signals from the single spot and raster scan experiments were integrated with respect to time, and the resulting charge-equivalent values were normalized to the detector output from the largest dose. The normalized data were then compared to that obtained from the ionization chamber, which was also normalized to the output to the largest dose, as shown in Fig. 3(a) for the single spot data and Fig. 3(b) for the raster scan data. This comparison is meaningful because both the GOPD and the ion chamber produce signals proportional to the collected charge generated by the interaction of proton radiation with the detector medium. In the GOPD, incident protons generate electron–hole pairs in $\beta$-$Ga_2O_3$, producing a transient current that is converted by the preamplifier to yield a voltage pulse. Consequently, the integrated voltage output corresponds to the total charge—and thus the dose—delivered during irradiation. Both the GOPD and the ion chamber showed a linear normalized output with respect to dose for both the single spot exposure and the raster scan, allowing for a one-to-one conversion for dosimetry purposes. This linear response of the GOPD with respect to dose is due to the increased proton dose generating more electron-hole pairs, with the relation given in (1) [35].

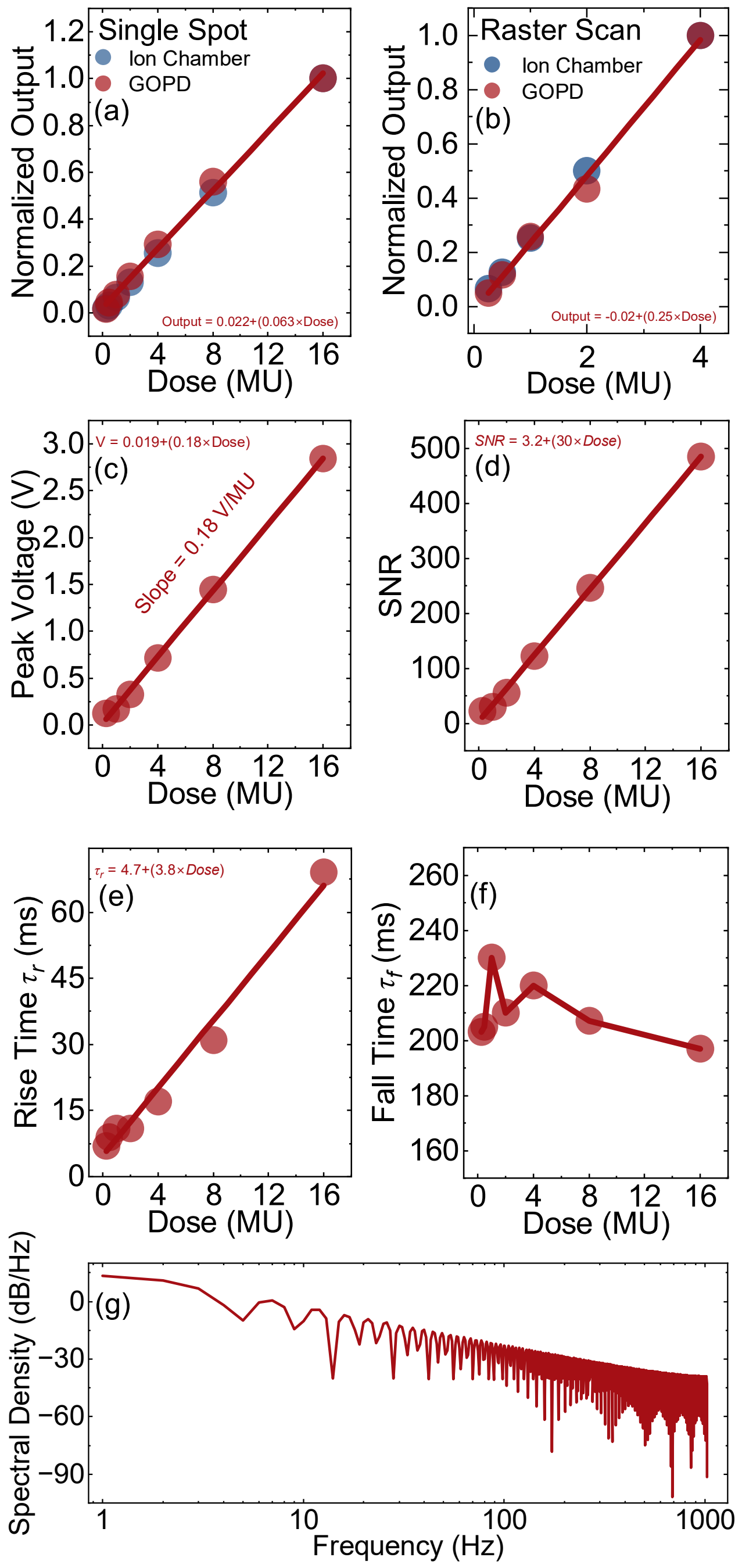

**Figure 3.** Integral of the voltage response of the GOPD to the (a) single spot and (b) raster scan proton radiation with respect to time compared to the charge response of the ion chamber, with both normalized to their max dose value. (c) Peak voltage of the signal with respect to dose. (d) SNR of the GOPD versus dose (e) $\tau_r$ and (f) $\tau_f$ of the GOPD from 0.26 MU to 16 MU. (g) GOPD's spectral density of the dark voltage.

$$Q = \Phi \times k \tag{1}$$

Where $\Phi$ is the proton fluence per unit area, $k$ is the average number of free electrons generated per proton at a specific energy, and $Q$ is the charge. The normalized output of the GOPD had a similar slope to the ion chamber, which shows similar changes in the detector's output with respect to dose, suggesting similar levels of sensitivity in this dose range.

Fig. 3(c) presents the peak voltage value from the GOPD for varying proton irradiation doses. A linear dependence between the GOPD's peak voltage amplitude and the delivered dose was observed. The data in Fig. 3(a) are similar to those in Fig. 3(c), as the proton flux increases with increasing dose, resulting in an increase in charge and peak current from the GOPD, and subsequently, an increase in voltage at the amplifier's output. This increase in the peak current with increasing dose resulted in the SNR of the detector increasing linearly with dose, as shown in Fig. 3(d).

The temporal characteristics of the detector response were characterized by analyzing $\tau_r$ and $\tau_f$ versus dose, which is presented in Figs. 3(e) and (f), respectively. The rise time, $\tau_r$, increased with dose, which could be due to the optimal dose rate of the MEVION accelerator being 0.394 MU/mS. The $\tau_f$ of the combined detector–amplifier system was on the order of hundreds of milliseconds, which is indicative of high defect densities within the material, consistent with prior reports[36].

The spectral density of the GOPDs dark voltage is displayed in Fig. 3(g), and it showed an approximately linear decrease with respect to frequency on the log-log plot. This linear decrease suggests that the primary source of dark noise is due to flicker noise [37, 38]. The flicker noise could be due to the front-end electronics, as well as poor electrical connections [38].

The GOPD was further evaluated under single-spot proton irradiation with energies ranging from 199 MeV to 60 MeV at a fixed dose of 2 MU. The transient voltage response of the detector at each energy level is presented in Fig. 4(a). A clear non-linear increase in the peak voltage with rising proton energy is demonstrated in Fig. 4(a). To quantitatively compare the responses, the voltage signals were integrated and normalized to the output corresponding to the highest energy. These normalized results were then compared to both the normalized response of the ion chamber and a simulated dose distribution incident on the GOPD, as shown in Fig. 4(b). The simulation was carried out using the RayStation treatment planning system, employing a Monte Carlo algorithm with a 1 mm spatial grid and an uncertainty threshold of 0.5%. Across the examined energy range, the GOPD response exhibited a consistent non-linear trend with both the simulation and the ion chamber measurements. The observed nonlinear reduction in both the transient response and the normalized output arises from the use of damping blocks in the MEVION accelerator to degrade the proton beam energy. Beam degradation through these spoiling elements increases proton scattering and leads to an enlarged beam radius. The resulting beam broadening reduces the proton fluence incident on the GOPD, thereby decreasing the measured signal. The proton fluence through the GOPD, calculated from the simulated dose, is presented in Fig. 4(c). As shown in Fig. 4(c), the fluence decreases with decreasing proton energy, which accounts for the reduced signal at lower energies, despite the increasing stopping power of the material over this energy range. The fluence was calculated with the relation given in (2).

$$\Phi = \frac{D(E)\rho}{S(E)} \tag{2}$$

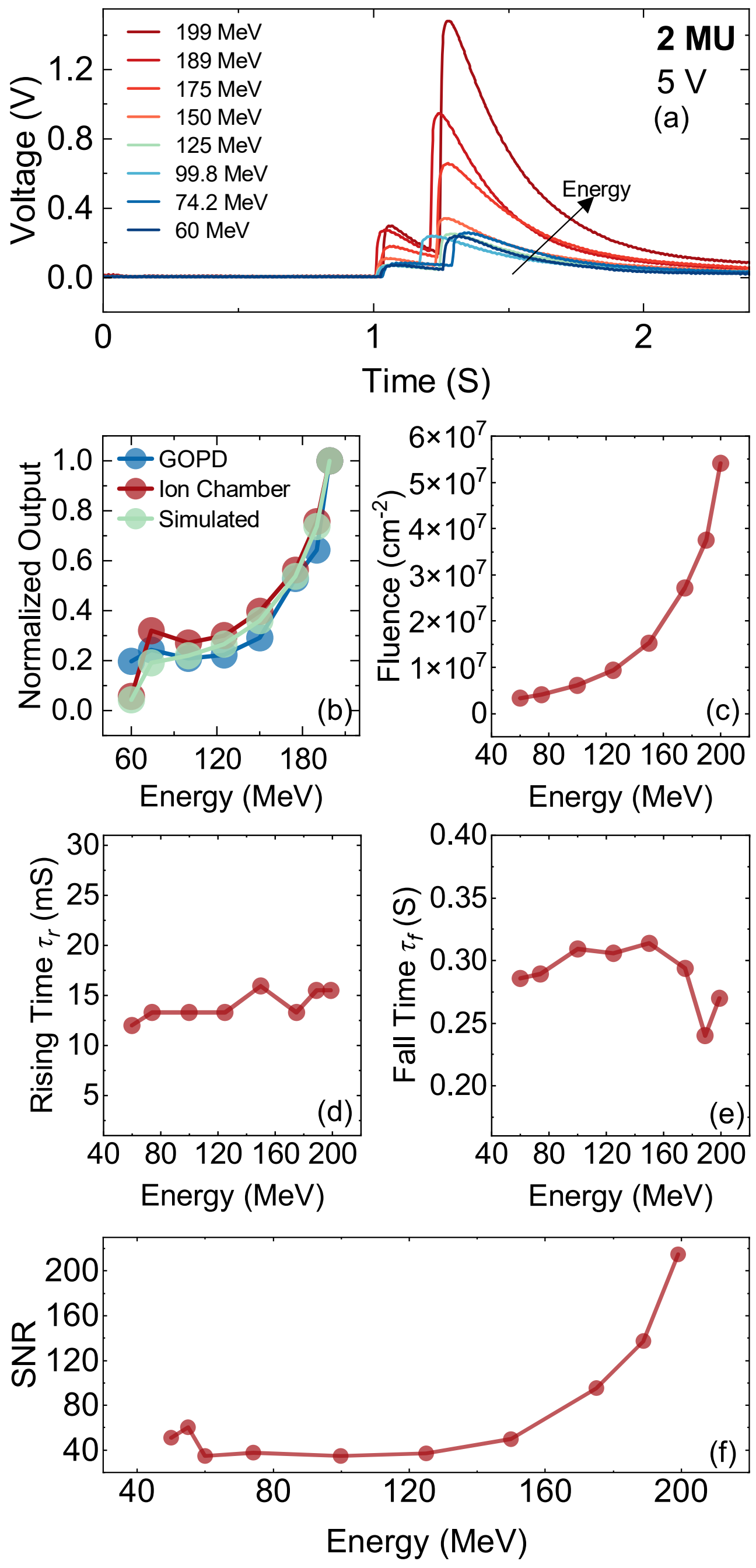

**Figure 4.** (a) Transient response of the GOPD to a single spot of proton radiation with a dose of 2 MU, and energies from 199 MeV to 60 MeV. (b) Integral of the voltage response of the GOPD with respect to time, the charge response of the ion chamber, and the simulated proton radiation dose that penetrated the GOPD, with all values being normalized to the value at 199 MeV. (c) Calculated proton fluence that passed through the GOPD. (d) $\tau_r$ and (e) fall $\tau_f$ for the transient voltage response of the GOPD from 60 MeV to 199 MeV. (f) SNR of the GOPD.

Where $\rho$ is the density of the GOPD, $D(E)$ is the simulated energy-dependent dose delivered to the GOPD, and $S(E)$ is the energy-dependent stopping power of the GOPD. An expression for $S(E)$ is given in (3).

$$S(E) = \frac{\mathrm{d}E}{\mathrm{d}x} \tag{3}$$

Where d$E$/d$x$ is the energy loss of the proton radiation per unit depth.

The energy dependence of $\tau_r$ is shown in Fig. 4(c), and $\tau_f$ is shown in Fig. 4(d). These figures show that $\tau_r$ and $\tau_f$ do not have a clear, significant energy dependence. The SNR of the detector with respect to energy is shown in Fig. 4(e), indicating that the SNR decreases with decreasing energy. This decrease is also related the the decreased fluence through the GOPD.

## IV. Conclusion

This work demonstrated the response characteristics of the GOPD across a wide energy range from 199 MeV to 60 MeV, incorporating a low-noise amplifier to enhance signal fidelity and maintain linearity with dose. The GOPD's measured response showed strong agreement with both ion chamber data and Monte Carlo simulations. Additionally, the detector effectively measured radiation doses ranging from 0.26 MU to 16 MU, with an integrated voltage response that closely paralleled those of the ion chamber. These results confirm the feasibility of β-$Ga_2O_3$ as a promising material for next-generation dosimetry devices in proton radiation therapy, addressing key limitations of current IMPT dosimetry technologies.

## AUTHOR DECLARATIONS

### Conflict of Interest

The authors have no conflicts to disclose.

## ACKNOWLEDGEMENT

The authors acknowledge the support from the Pilot Funding through the HCI/Engineering Innovation in Cancer Engineering (ICE) Partnership Seed Grants, the University of Utah start-up fund, and PIVOT Energy Accelerator Grant U-7352FuEnergyAccelerator2023. This work was performed in part at the Utah Nanofab Cleanroom sponsored by the John and Marcia Price College of Engineering College of Engineering and the Office of the Vice President for Research. The authors appreciate the support of the staff and facilities that made this work possible.

## DATA AVAILABILITY

The data that support the findings of this study are available from the corresponding authors upon reasonable request.